# *Self-driven beta-radiation detection through a graphene/water/silicon heterostructure*

*Tengfei Wu[1], Kangchen Xiong[2], Minhui Yang[2], Xin Chen[2], Shisheng Lin[1,2,3]**

[1] College of Integrated Circuits, Zhejiang University, Hangzhou, China.

[2] College of Information Science and Electronic Engineering, Zhejiang University, Hangzhou, China.

[3] State Key Laboratory of Extreme Photonics and Instrumentation, Zhejiang University, Hangzhou, China.

*Corresponding author. Email: shishenglin@zju.edu.cn

## Abstract

Large-volume neutrino observatories, exemplified by IceCube, exploit Cherenkov light generated by neutrino-induced charged particles in Antarctic ice, highlighting the potential of water-based media for high-energy particle detection. In β-decay, a neutron converts into a proton via the weak interaction, emitting an electron and an electron antineutrino. The electron constitutes the beta radiation detected in this work, while the accompanying antineutrino motivates future extensions toward neutrino detection. Here we demonstrate a complementary electrical transduction strategy using a graphene/water/silicon heterostructure that generates reversible current transients under beta irradiation at zero applied bias. Energetic electrons deposit energy in silicon, producing nonequilibrium carriers that perturb the semiconductor surface potential and electric displacement across the adjacent aqueous interface. The resulting redistribution of interfacial water polarization drives a transient compensation current, whereas carrier relaxation restores the initial polarization state and generates an opposite-polarity recovery signal. The complete irradiation-detector-electrometer system exhibits exponential rise and recovery time constants of approximately 1.0 and 3.6 ms, respectively. The response is strongly regulated by water-layer thickness and ionic composition, and the 17 beam pulses of a programmed 1 s sequence at 17 Hz are resolved individually. Integration with Ce:YAG further introduces a scintillation-mediated excitation pathway. These findings establish interfacial water polarization as a tunable, zero-bias electrical transduction mechanism for time-resolved beta-radiation monitoring and motivate future investigations of higher-energy charged particles and secondary particles relevant to neutrino detection.

## Introduction

Radiation detectors underpin dosimetry, irradiation control, nuclear monitoring and radiation-safety systems.[1-3] Beta radiation is commonly detected using gas ionization, scintillation or semiconductor charge collection, each of which converts deposited radiation energy into an electrical signal through a distinct physical pathway.[4,5] Gas-filled instruments provide robust counting but generally require high-voltage operation and sealed or gas-handling components. Scintillation systems introduce an optical conversion and photodetection chain, whereas semiconductor detectors usually rely on built-in or externally applied electric fields to collect radiation-generated carriers.[6-10] A compact architecture that reports irradiation timing directly at zero external detector bias could therefore complement established platforms for continuous monitoring.[11-15]

Graphene provides a sensitive electrical interface because its electronic state is strongly modulated by electrostatic charge, electrolyte gating and changes in interfacial potential.[16-18] At charged solid–water interfaces, the structure and polarization of interfacial water are coupled to surface charge and the local electric field.[19-21] At graphene-aqueous interfaces, electrical-double-layer formation, ionic screening and substrate coupling further regulate interfacial charge organization.[22-28] In parallel, dynamic water-semiconductor junctions have shown that the polarization state of interfacial water can couple mechanical or optical perturbations to direct electrical output.[29,30] Graphene/water/silicon photodetectors further indicate that semiconductor-generated carriers can modulate the interfacial field and produce transient current through polarization and depolarization of the aqueous layer.[31,32]

Here we extend this interfacial-polarization framework to time-resolved beta-radiation monitoring in a graphene/water/silicon stack. Following the experimental sequence in Figs. 1-6, we first establish the device architecture and energetic-electron interaction with silicon, then describe the reversible interfacial polarization cycle, quantify the transient timescale, examine regulation by liquid geometry and ionic composition, demonstrate event-resolved tracking of a programmed 17 Hz irradiation sequence, and finally test a Ce:YAG-coupled hybrid configuration.

## Results

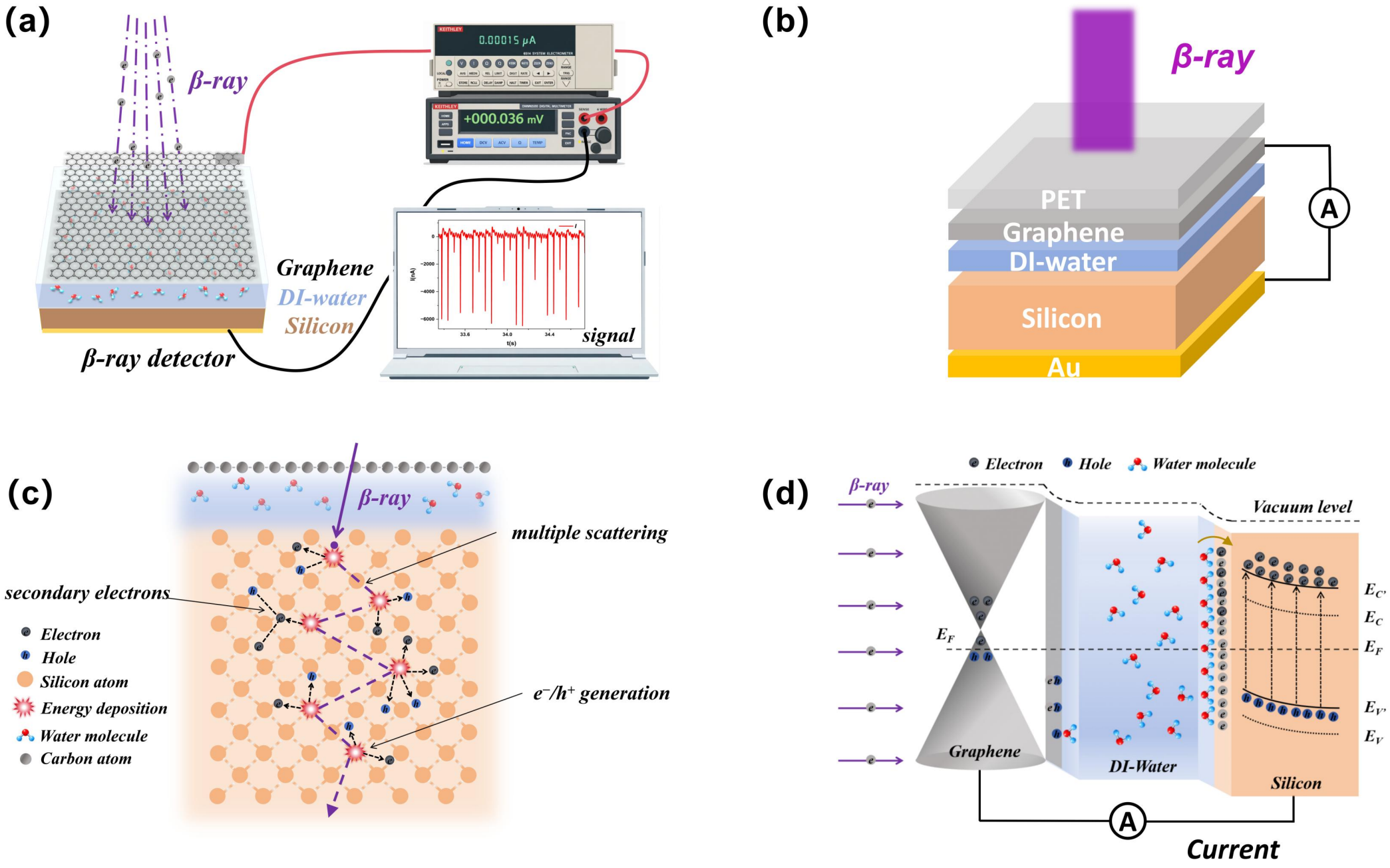

**Fig. 1 | Device architecture and beta-to-interface transduction.** a, Experimental configuration of the graphene/water/silicon detector connected to an electrometer for zero-bias current readout, with the transient signal acquired in real time. b, Cross-sectional device structure comprising a PET-supported graphene, a deionized-water interlayer, silicon and a rear Au contact. c, Schematic of energetic-electron interaction with silicon through multiple scattering, secondary-electron generation and local energy deposition. d, Energy-band diagram of the graphene/water/silicon stack under beta irradiation, showing carrier generation in silicon and the resulting perturbation of the interfacial electrostatic state that drives a transient current in the external circuit.

The detector is built as a vertical graphene/water/silicon heterostructure with a rear Au electrode and a PET-supported graphene top electrode (Fig. 1a,b). The graphene and silicon contacts are connected directly to a Keithley 6514 electrometer. All radiation response measurements reported here were performed without an externally applied detector bias.

Energetic electrons entering the active region undergo multiple scattering and inelastic energy loss, generating secondary excitations and electron-hole populations in silicon (Fig. 1c). These nonequilibrium carriers modify the electrostatic state near the silicon/water boundary. Because the aqueous layer is electrically coupled to both silicon and graphene, the semiconductor perturbation changes the interfacial electric displacement and drives a transient compensation current in the external circuit (Fig. 1d). The device therefore separates radiation interaction from signal transduction: silicon provides the principal solid interaction volume, whereas the liquid interface converts the resulting electrostatic perturbation into an externally measurable current.

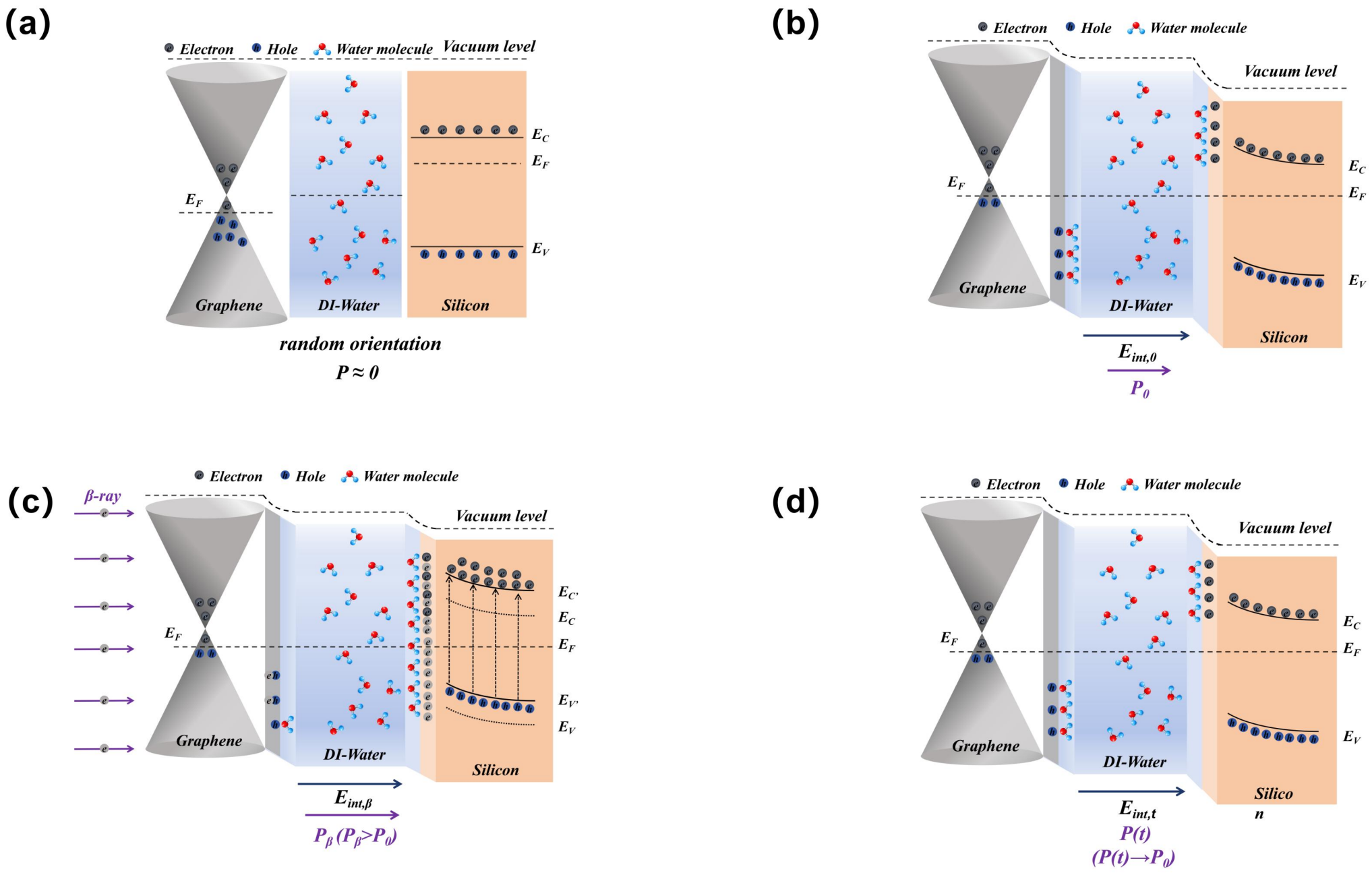

**Fig. 2 | Reversible interfacial polarization during beta irradiation.** a, Before equilibration, graphene, water and silicon retain independent electronic states and the water layer shows random molecular orientation with negligible net polarization. b, Contact-induced equilibration aligns the energy bands and establishes the dark interfacial field $E_{int,0}$ and polarization $P_0$. c, Radiation-generated carriers in silicon shift the interfacial field toward $E_{int,\beta}$ and the polarization toward $P_\beta$. d, Carrier relaxation drives the interfacial polarization back toward the dark state, $P(t) \rightarrow P_0$.

Figure 2 resolves the proposed response into a reversible interfacial cycle. Before equilibration, graphene, water and silicon retain their individual electronic states and the aqueous layer has little net orientational polarization (Fig. 2a). After assembly, charge redistribution establishes a dark interfacial field $E_{int,0}$ and a corresponding polarization state $P_0$ (Fig. 2b). Here, $P_0$ denotes an effective interfacial polarization rather than uniform molecular alignment throughout the full liquid layer.

Under beta irradiation, carrier generation in silicon changes the surface potential and shifts the interface toward an irradiation-modified field $E_{int,\beta}$ and polarization $P_\beta$ (Fig. 2c). The external circuit supplies or removes compensation charge while the electric displacement evolves. When the carrier perturbation relaxes, the interface returns toward the dark state and the direction of the polarization change reverses (Fig. 2d). The transient current follows

$$\boldsymbol{D} = \boldsymbol{\varepsilon_0 E} + \boldsymbol{P} \qquad (1)$$

$$\boldsymbol{I_{\mathrm{ext}}} = \boldsymbol{A}\frac{\boldsymbol{dD}}{\boldsymbol{dt}} = \boldsymbol{A\varepsilon_0}\frac{\boldsymbol{dE}}{\boldsymbol{dt}} + \boldsymbol{A}\frac{\boldsymbol{dP}}{\boldsymbol{dt}} \qquad (2)$$

The signal is therefore largest while the interfacial electrostatic state is changing, which naturally produces a transient response and an opposite-sign recovery component.

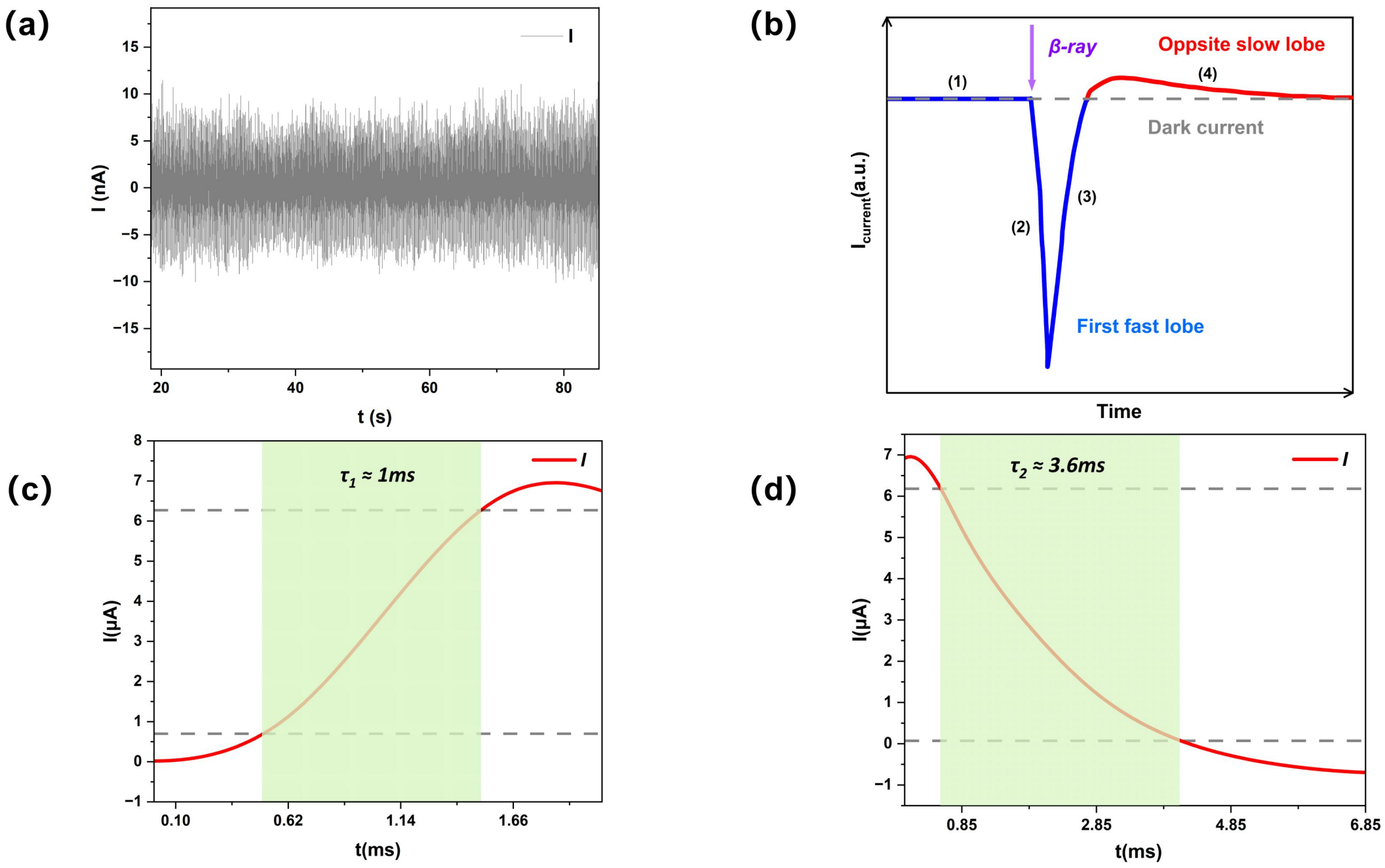

**Fig. 3 | Dark current, bipolar transient and millisecond dynamics.** a, Dark-current trace recorded without irradiation. b, Characteristic bipolar response, indicating the dark baseline (1), the first fast lobe (2, 3) and the opposite-sign slow recovery lobe (4). c, Single-event rising phase fitted with an exponential time constant of approximately 1.0 ms. d, Recovery phase fitted with an exponential time constant of approximately 3.6 ms.

The dark electrical trace exhibits a fluctuating baseline in the absence of irradiation (Fig. 3a). A characteristic irradiation event is bipolar: a fast first lobe is followed by an opposite-sign, slower recovery component (Fig. 3b). The paired waveform is consistent with rapid displacement of the interface away from its dark equilibrium followed by relaxation back toward the initial polarization state.

The rising and recovery phases of single-event transients are well described by exponential relaxation, with time constants of approximately 1.0 ms and 3.6 ms, respectively (Fig. 3c,d). Because the beam pulse width of several tens of microseconds is much shorter than both constants, these values characterize the relaxation of the detector-electrometer system following each pulse rather than the temporal profile of the beam; the electrometer bandwidth and the online sampling rate further limit the resolution of faster components, so the quoted constants represent upper bounds on the intrinsic interfacial relaxation. The recovery time is far shorter than the 58.8ms period of a 17Hz irradiation sequence, providing temporal separation between successive programmed events.

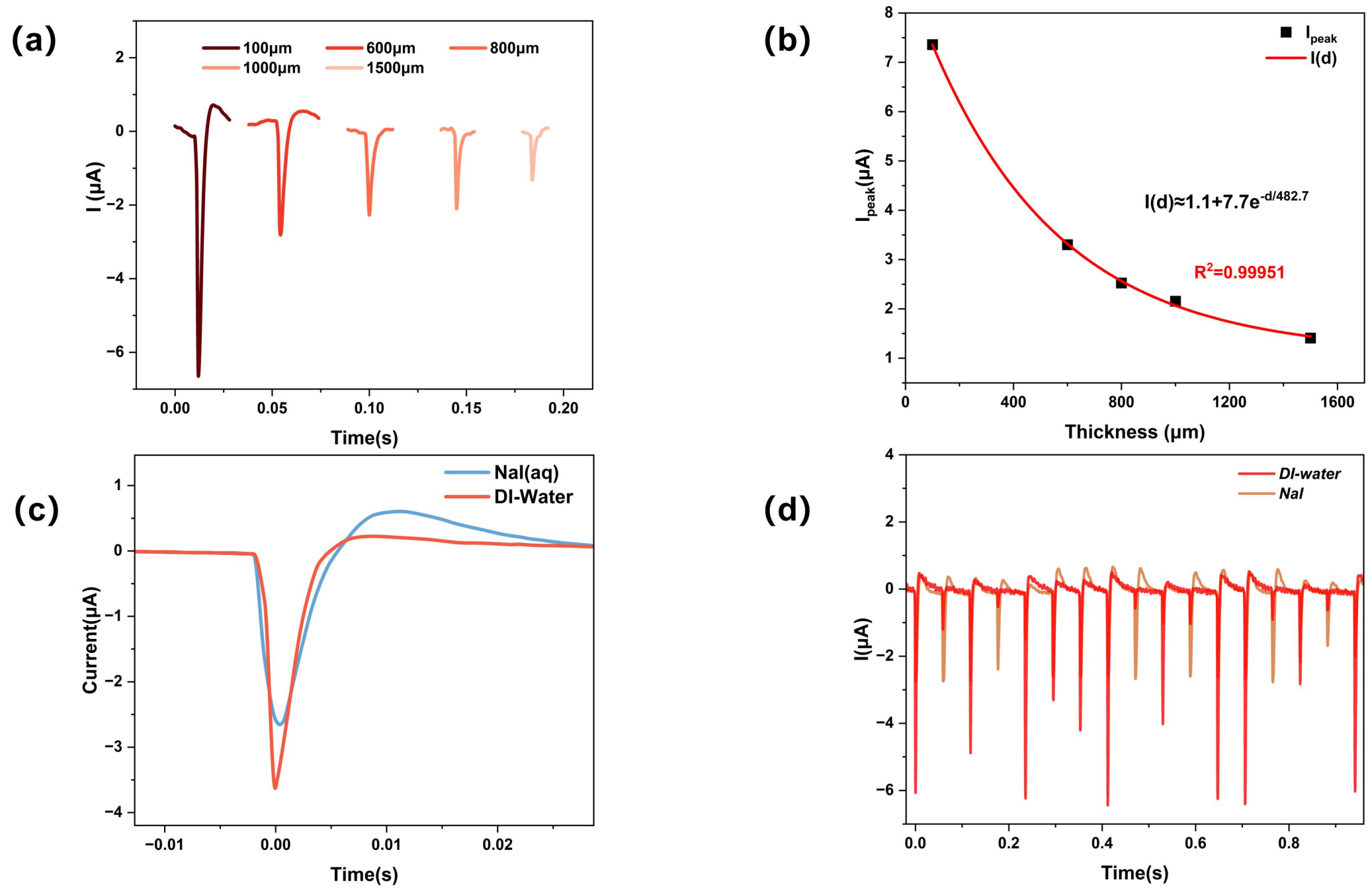

**Fig. 4 | Liquid-layer regulation of the beta-radiation response.** a, Transient responses for nominal water-layer thicknesses from 100 to 1,500 μm. b, Extracted peak current as a function of water-layer thickness, fitted with an exponential decay. c, Single-event transient responses of deionized water and 0.15 mol $l^{-1}$ NaI solution. d, Repeated transient trains recorded from deionized water and 0.15 mol $l^{-1}$ NaI solution.

The aqueous layer provides a direct handle for tuning the detector response. Varying the nominal water thickness from 100 to 1,500 μm changes both the amplitude and temporal profile of the transient waveform (Fig. 4a,b).

Aqueous composition provides an independent tuning parameter. Replacing deionized water with 0.15 mol $l^{-1}$ NaI reduces the amplitude of the first lobe of the single-event response and enhances the opposite-polarity recovery component (Fig. 4c), while the repeated transient train retains the programmed event timing at a reduced amplitude (Fig. 4d). The NaI response may reflect changes in ionic screening, interfacial charge organization and dielectric relaxation.Together, the thickness and NaI measurements identify the liquid as an active transduction layer whose geometry and chemistry influence signal formation.

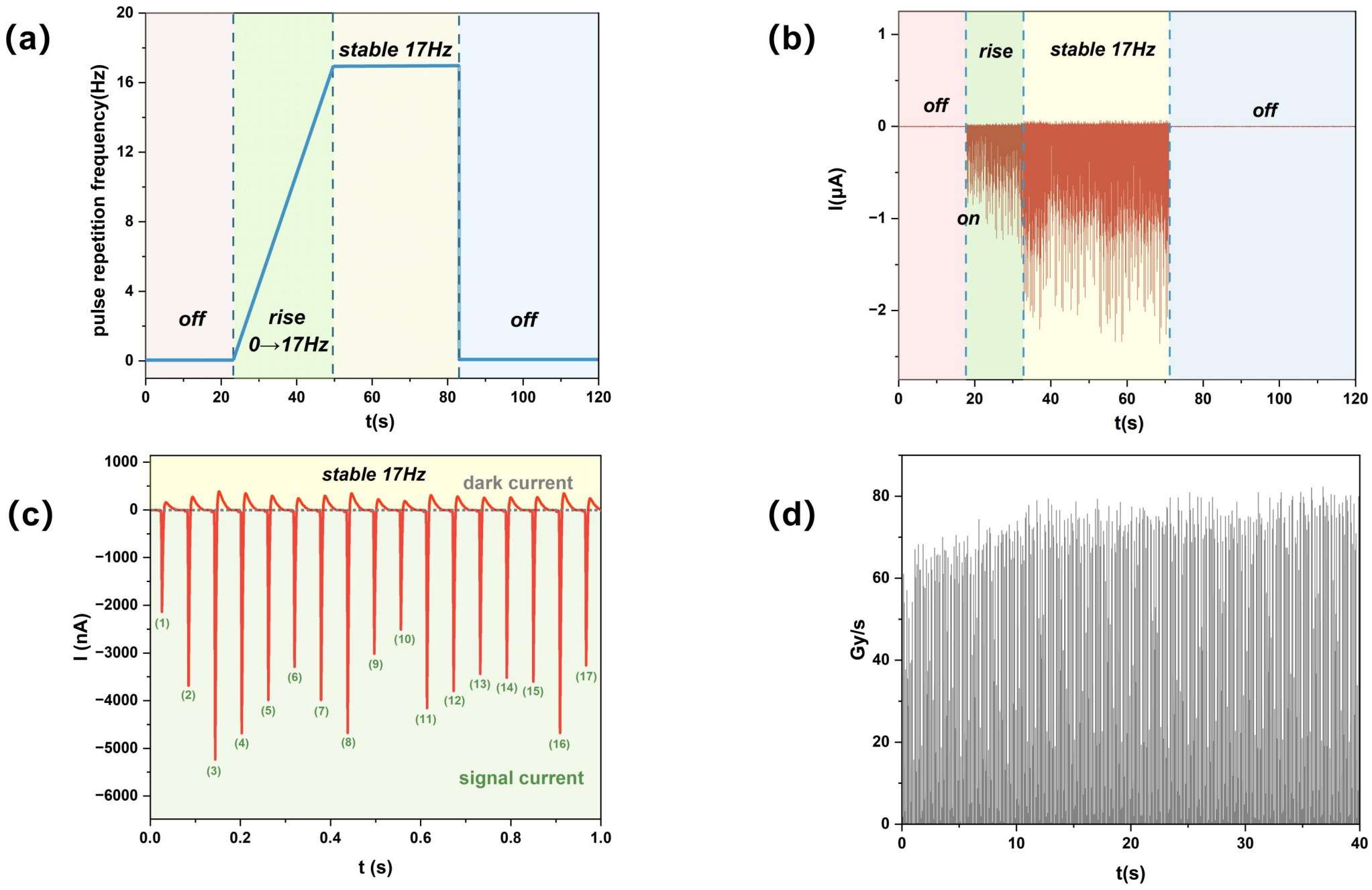

**Fig. 5 | Zero-bias monitoring of a programmed 17 Hz irradiation sequence.** a, Facility protocol comprising pulse-off, 0-17 Hz ramp, stable 17 Hz and final pulse-off segments. b, Detector current recorded over the full irradiation protocol, with the irradiation stages marked. c, One-second interval during stable operation containing 17 resolved electrical events, numbered (1)-(17). d, Facility dose-rate record acquired during irradiation.

We next tested whether the detector could reproduce the temporal structure of a programmed irradiation sequence over an extended interval. The facility protocol contains an initial off state, a ramp from 0 to 17 Hz, a stable 17Hz interval and a final off state (Fig. 5a). The zero bias detector current follows these stages directly (Fig. 5b): discrete transients appear during the frequency ramp, become dense and repetitive during stable irradiation, and disappear when the pulse train is disabled.

A one-second segment from the stable interval contains 17 resolved current events, numbered (1)-(17), matching the programmed 17Hz repetition frequency (Fig. 5c). Within this selected interval, the event count is consistent with the detector preserving timing information in the irradiation sequence rather than reporting only integrated dose. The facility dose-rate record provides an independent irradiation reference (Fig. 5d), and separate silver-dichromate chemical dosimetry verified kGy-level accumulated exposure.

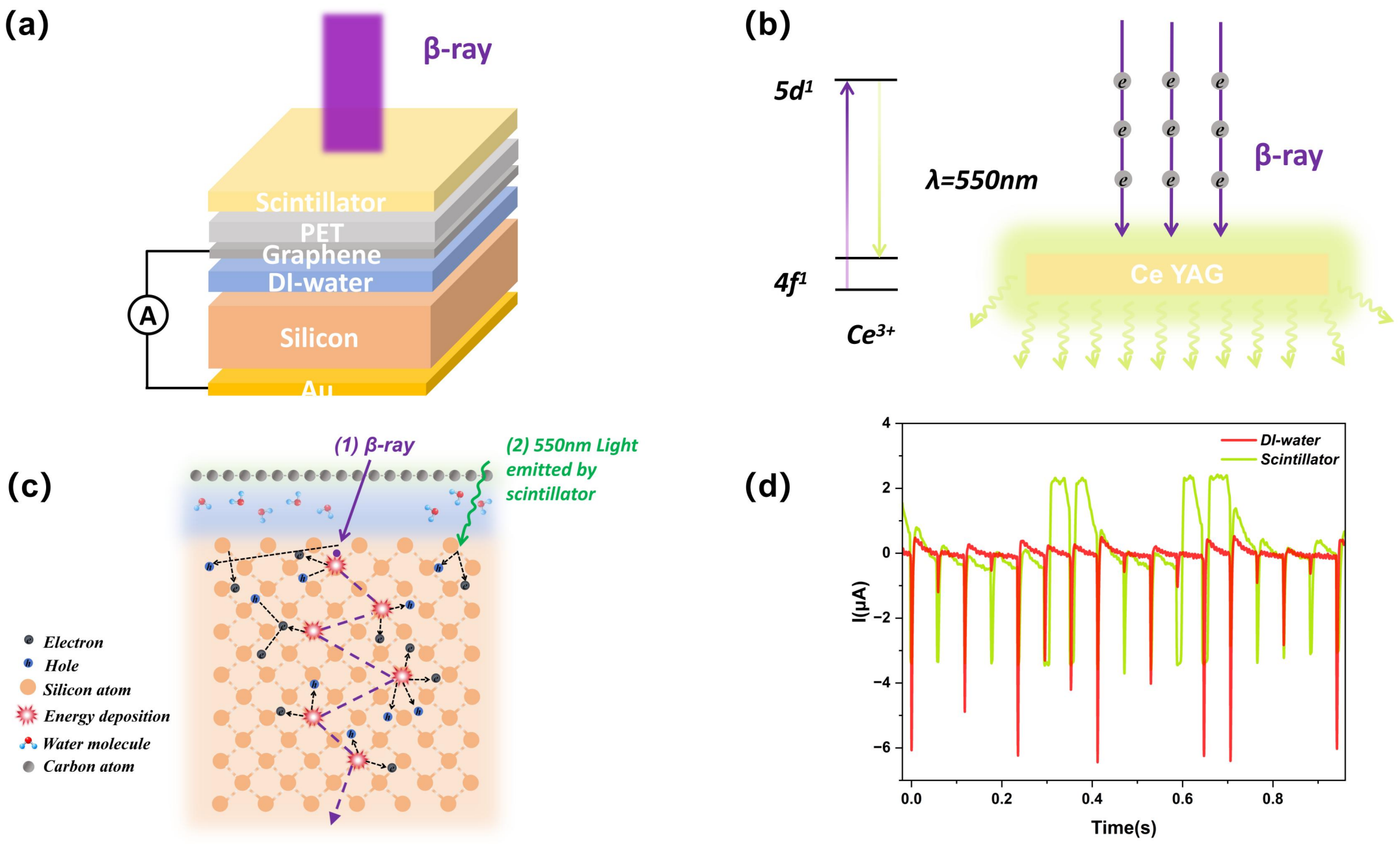

**Fig. 6 | Direct and scintillator-coupled excitation of the liquid-interfaced transducer.** a, Ce:YAG scintillator integrated above the graphene/water/silicon detector. b, Simplified $Ce^{3+}$ scintillation scheme with green emission near 550 nm following energetic-electron excitation. c, The two excitation pathways: (1) direct energetic-electron deposition in silicon and (2) scintillation-mediated optical excitation, both generating nonequilibrium carriers that perturb the liquid-coupled interface. d, Measured comparison between the direct aqueous detector and the Ce:YAG-coupled configuration.

The liquid-interfaced transducer can also be combined with an upstream optical-conversion layer. In the hybrid configuration, a Ce:YAG scintillator is placed above the graphene/water/silicon stack (Fig. 6a). Energetic electrons excite $Ce^{3+}$ centres and generate green emission near 550 nm (Fig. 6b).[33–37] Photons absorbed by silicon create carriers that can perturb the same liquid-coupled interface involved in direct electron excitation (Fig. 6c). The direct and scintillator-coupled configurations show clearly different temporal signatures (Fig. 6d). The direct device retains relatively sharp repeated transients, whereas the Ce:YAG-coupled stack exhibits broader components and baseline excursions superimposed on the repeated response. Because energetic electrons may also penetrate or scatter through the scintillator, the hybrid waveform represents the response of the complete electron-scintillator-detector stack rather than a pure isolated optical signal. Nevertheless, the experiment demonstrates that the same liquid-interfaced electrical transducer is compatible with both direct energetic-electron excitation and an added scintillation-mediated pathway.

## Discussion

The measurements establish a zero-bias liquid-interfaced architecture for time-resolved beta-radiation monitoring. Energetic electrons deposit energy in silicon and generate a transient carrier population that displaces the graphene/water/silicon stack from its dark electrostatic equilibrium. The aqueous interface converts this perturbation into a time-dependent electric-displacement current, while subsequent relaxation generates the opposite-sign recovery lobe and restores the baseline. In this picture, the liquid is not a passive spacer but a functional electrical transduction layer.

The sequence of experiments provides several internally consistent signatures of this mechanism. The reversible polarization picture accounts for the bipolar response; water-layer thickness strongly changes the measured waveform; dissolved NaI reshapes the bipolar waveform and reduces the response amplitude; and the millisecond recovery permits event-resolved detection of a 17 Hz sequence.

The main functional advance is the preservation of irradiation timing at zero external detector bias. The one-second 17 Hz trace demonstrates that the architecture can resolve a programmed pulse sequence without a scintillator or photodetector in the direct configuration. The Ce:YAG experiment then shows that an optical conversion layer can be incorporated when desired without changing the underlying liquid-interfaced electrical readout concept.

More broadly, the ability to tune the response through liquid geometry and composition introduces a design degree of freedom that is difficult to access in fixed solid-solid junctions. This combination of semiconductor radiation interaction and chemically configurable liquid-interface transduction may provide a route toward compact self-driven monitors for pulsed irradiation environments.

# Methods

## Device assembly

The detector active area was 1 cm × 1 cm. Unless otherwise stated, the aqueous layer was approximately 100 μm thick; thickness-dependent measurements used nominal layers from 100 to 1,500 μm. The liquid was either deionized water or 0.15 mol $l^{-1}$ NaI solution, as specified for each measurement. The device architecture comprised a PET-supported graphene electrode, an Ag contact to graphene, the aqueous interlayer, a silicon substrate and an Au rear contact.

## Electrical characterization

Current measurements were performed with a Keithley 6514 electrometer. Radiation-response measurements were carried out at zero externally applied detector bias. Online-monitoring traces were sampled at 1 kHz. Current polarity follows the fixed electrometer lead convention; reversing the measurement leads reverses the displayed sign without changing the transient magnitude.

## Irradiation protocol and dosimetry

Beta irradiation was provided by a pulsed electron beam from a 10 MeV linear electron accelerator at an industrial irradiation facility. The accelerated electrons are physically identical to beta-minus particles and are hereafter referred to as beta radiation. The source-to-detector distance was 1.3 m, the electron energy was 10 MeV, and the beam pulse width was several tens of microseconds. The repetition-frequency protocol contained an initial pulse-off interval, a ramp from 0 to 17 Hz, a stable 17-Hz interval and a final pulse-off interval. During off stages the pulse generator was disabled while the detector geometry was unchanged. The facility dose-rate monitoring point and the silver-dichromate chemical dosimeters were positioned in the same irradiation plane as the detector, so that the recorded dose rate and the accumulated exposure both refer to the detector location. Absorbed exposure was checked independently using commercial silver-dichromate liquid chemical dosimeters.[38–40] The facility dose-rate record and chemical dosimetry were used as independent irradiation references for the online detector measurements.

**Scintillator-coupled control**

For the scintillator-coupled measurement, a Ce:YAG scintillator was introduced above the graphene/water/silicon stack as illustrated in Fig. 6. The figure uses a simplified $Ce^{3+}$ emission scheme with an optical band centred near 550 nm to represent the scintillation-mediated excitation route. Scintillation photons absorbed in silicon generate carriers that perturb the same interfacial electric field involved in direct energetic-electron excitation. The measured electrical trace therefore represents the response of the complete scintillator-detector-interface-electrometer configuration.

# Acknowledgements

S.L. acknowledges support from the National Natural Science Foundation of China (grant nos. 51202216, 51551203, 61774135 and 62474161) and the Distinguished Youth Fund of the Zhejiang Provincial Natural Science Foundation (LR21F040001).

## Author contributions

S.L. conceived the study, designed the experiments, coordinated the overall experimental work, and wrote the paper. T.W. carried out the experiments, wrote the paper and discussed the results. K.X., M.Y. and X.C. discussed the results . All authors contributed to the preparation of the manuscript.

## Competing interests

The authors declare that they have no competing interests.

## Data availability

The data that support the findings of this study are available from the corresponding author upon reasonable request.

## Correspondence

Correspondence and requests for materials should be addressed to Shisheng Lin (shishenglin@zju.edu.cn).